\documentclass[amssymb,amsmath,aip,jcp,preprint]{revtex4-1}
\usepackage{bm}%
\usepackage[colorlinks=true,linkcolor=blue]{hyperref}%
\usepackage{graphicx}
\usepackage{epstopdf}
\usepackage{color,soul}
\begin{document}
\title{Non-linear non-local molecular electrodynamics with nano-optical fields}

\author{Vladimir Y.  \surname{Chernyak}}
\email{chernyak@chem.wayne.edu}

\affiliation{Department of Chemistry, Wayne
State University, 5101 Cass Ave, Detroit, MI 48202}
\affiliation{Department of Mathematics, Wayne
State University, 656 W. Kirby, Detroit, MI 48202}
\author{Prasoon \surname{Saurabh}}
\email{psaurabh@uci.edu}
\author {Shaul \surname{Mukamel}}
\email{smukamel@uci.edu}
\affiliation{Department of Chemistry, University
of California, Irvine, CA 92697}

\date{Compiled \today}

\begin{abstract}
The interaction of optical fields sculpted on the nano-scale with matter may not be described by the dipole approximation since the fields vary appreciably across the molecular length scale. Rather than incrementally adding higher multipoles it is advantageous and more physically transparent to describe the optical process using non-local response functions that intrinsically include all multipoles. We present a semi-classical approach to the non-linear response functions based on the minimal coupling Hamiltonian. The first, second and third  order non-local response functions are expressed in terms  of correlation functions of the charge and the current densities. This approach is based on the gauge invariant current rather than the polarization, and on the vector potential rather than the electric and magnetic fields.
\end{abstract}

\maketitle

\section{Introduction}

Quantum optics is commonly formulated by making the semiclassical approximation whereby the fields are treated classically and matter is treated quantum mechanically \cite{mukamel1995,scully1997,cohen1997photons}. The radiation-matter coupling may be described by the minimal coupling Hamiltonian where the electromagnetic fields are represented by vector potential $\bm{A}(\bm{r},t)$ \citep{aharonov1961} and the matter properties enter through the current $\hat{\bm{j}}(\bm{r},t) $ and charge density $\hat{\sigma}(\bm{r},t)$ matrix elements of the desired transitions \cite{rammer2004quantum}. Alternatively, the multipolar hamiltonian is used where the electromagnetic field is represented by the electric and magnetic fields and matter is expanded in electric and magnetic multipoles. In most applications, including the theory of the laser \cite{scully1997}, the lowest order of this multipolar expansion, known as the dipole (long wavelength) approximation is sufficient to account for experimental observations \cite{feynman1969}.

In recent years, however, there has been rapid developments \cite{bharadwaj2009optical,stockman2011nanoplasmonics,stockman2015quantum,donehue2014plasmon,kalashnikov2014quantum}, both theoretical and experimental, in the field of nanooptics \cite{novotny2012principles}. Notable for spectroscopy are, nanoantenna \cite{bharadwaj2009optical}, nanoplasmonic \cite{stockman2011nanoplasmonics,stockman2015quantum,ding2014quantum} and associated spatial and temporal resolutions of optical spectra resulting in attosecond local-field enhanced ($\times \;10-10^2$) spectroscopy \cite{stockle2000nanoscale,bailo2008tip,brongersma2007surface,kuhn2006enhancement,brown2015fan}. This is possible because of the nano-scale field confinement ($ \sim 10nm$) \cite{bharadwaj2009optical}, as in case of nanoantenne, and spectral bandwith of plasmonic spectra ($ 850-2200\; THz$) \cite{stockman2011nanoplasmonics}. $\textit{Nano-optical fields}$ are spatially confined fields which show appreciable  position ($\bm r$)- dependence on the molecular length scale \cite{minovich2011generation,aeschlimann2007adaptive,novotny2012principles,abadeer2014distance,zhang2013}. Combination of a nanoantenna in tip of the probe \cite{novotny2012principles,taminiau2008optical} and utilizing their nanoplasmonic properties has pushed the boundaries of spectral resolution of optical spectroscopy, for example, surface- and tip-enhanced absorptions and Raman spectroscopy upto single molecule level \cite{kuhn2006enhancement,stockle2000nanoscale,bailo2008tip,donehue2014plasmon}. Non-linear optical manipulations of nano-particles have been reported as well \cite{kudo2012proposed,palomba2009near}. When the field confinement is comparable to quantum confinement of molecular orbitals, the dipole approximation may not be adequate and higher multipoles are required \cite{myroshnychenko2008modelling,bharadwaj2009optical}. 

In this paper, we calculate matter/field energy exchange and heterodyne detected optical signals [Eq.~\ref{absorptionrate}, Eq.~(\ref{S-het-explicit})] using non-local response functions that take all multipoles into account. Using the minimal coupling Hamiltonian for the interaction of nanoscale confined optical fields with molecules we can fully describe the radiation by a single field (the vector potential or the electric field or the magnetic field) and we do not need to use both the electric with magnetic fields. Starting with minimal coupling Hamiltonian [Eq.~(\ref{ht})] \cite{mukamel1995,rzk04,tanaka2001time}, we write the radiation field Hamiltonian in second quantized modes of optical field [Eq.~(\ref{hr})] and radiation-matter interaction Hamiltonian  using gauge-invariant current density $\hat{\bm{J}}(\bm{r},t)$ [Eq.~(\ref{int1})] \cite{tanaka2001time}. The material properties enter via correlation functions of the current density operator $\hat{\bm{j}}(\bm{r}) $  and the charge density operators $\hat{\sigma}(\bm{r})$ [Eqs.~(\ref{jr}) and (\ref{rho})].

The non-local response approach is particularly suitable for describing the interaction with nano-optical fields \cite{novotny2012principles,stockman2011nanoplasmonics}. Low order multipoles are adequate when the field varies slowly on the molecular scale. Which is not the case in nanooptics \cite{bharadwaj2009optical}. Coupling the microscopic Schr\"odinger equation with the macroscopic Maxwell’s equations \cite{salam2010molecular,romero2002quantum} then is simply an approximate. The non-local response provides a natural exact link for computing optical signals.

We work with gauge-dependent quantities, such as the vector $\bm{A}(\bm{r},t)$ and the scalar potentials $A_{0}(\bm{r},t)$, making sure that all observables, are gauge invariant. Gauge-invariant systems are most naturally formulated within the Lagrange formalism that involves path integrals. The Hamiltonian formalism is a bit more tricky, since gauge invariance leads to  the constraint $\bm{\nabla}\cdot \bm{E}- 4\pi\sigma= 0$, and the gauge-invariant formalism requires restricting the space of states to the physical subspace on which the constraints are satisfied. Alternatively one can fix the gauge by imposing a set of conditions on the potentials. The most suitable gauge for doing quantum calculations within the Hamilton approach is known as the Coulomb (or sometimes Hamiltonian) gauge, and is defined by the condition $\bm{\nabla}\cdot \bm{A}(\bm{r},t)= 0$. In this gauge the field variables are represented by a transverse vector potential, the longitudinal field is represented by the Coulomb potential created by the charges and therefore becomes a material, rather than field variable, whereas the total Hamiltonian $H$ of a system interacting with the electromagnetic field is given by \cite{mukamel1995,tanaka2001time},
\begin{eqnarray}
\label{ht}
\hat{H} &=& \hat{H}_{0}+ \hat{H}_{{\rm rad}}+ \hat{H}_{{\rm int}}= \hat{H}_{{\rm m}}+ \hat{H}_{{\rm c}}+ \hat{H}_{{\rm rad}}= \hat{H}_{{\rm m}}+ \hat{H}_{{\rm f}}, \nonumber \\
\hat{H}_{{\rm m}} &=& \hat{H}_{0}- \hat{H}_{{\rm c}}+ \hat{H}_{{\rm int}}, \;\;\; \hat{H}_{{\rm f}}= H_{{\rm rad}}+ \hat{H}_{{\rm c}},
\end{eqnarray}
where $\hat{H}_{0}$ is the material Hamiltonian that includes the kinetic energy of the charges, as well as the Coulomb interaction between them, $\hat{H}_{{\rm rad}}$ represents the (transverse) radiation field and can be written as Eq.~(\ref{hr}) in second quantized form as \cite{tanaka2001time,fetter2003quantum},

\begin{eqnarray}
\label{hr}
\hat{H}_{{\rm rad}}&=& \sum_{\bm{q} \lambda}\hbar \omega_{\bm{q} \lambda}\hat{b}^{\dagger}_{\bm{q}\lambda}\hat{b}_{\bm{q}\lambda}\nonumber\\&=& \frac{1}{8\pi}\int d\bm{r}\left(\hat{\bm{E}}_{{\rm T}}^{2}(\bm{r})+ (\bm{\nabla}\times \hat{\bm{A}}(\bm{r}))^2\right),
\end{eqnarray}
with,
\begin{eqnarray}
&[\hat{b}_{\bm{q}\lambda},\hat{b}_{\bm{q}^{\prime}\lambda^{\prime}}^{\dagger}] =  \delta_{\bm{q},\bm{q}^{\prime}}\delta_{\lambda,\lambda^{\prime}}.
\end{eqnarray}
Where $b_{\bm{q}\lambda}$ and $b^{\dagger}_{\bm{q}\lambda}$ are the annihilation and creation operators, respectively for the photon modes with wavevector $\bm{q}$ and polarization $\lambda$, while $\hat{\bm{E}}_{{\rm T}}(\bm{r})$ is the operator of the transverse electric field; $H_{{\rm int}}$ is the radiation-matter interaction Hamiltonian \cite{tanaka2001time},
\begin{eqnarray}
\label{int} \hat{H}_{{\rm int}}= -\int d\bm{r}\left(\hat{\bm{j}}(\bm{r})\cdot \hat{\bm{A}}(\bm{r}) -\frac{e^2}{2mc}\hat{\sigma}(\bm{r})\hat{\bm{A}}^{2}(\bm{r})\right).
\end{eqnarray}
This can be recast as,
\begin{eqnarray}
\label{define-H-int} \hat{H}_{{\rm int}}=-\int d\bm{r}\hat{\bm{J}}_{{\rm int}}(\bm{r})\cdot \hat{\bm{A}}(\bm{r}),
\end{eqnarray}
where,
\begin{equation}
\label{define-J-int}
\hat{\bm{J}}_{{\rm int}}(\bm{r})=\hat{\bm{j}}(\bm{r})-\frac{e^{2}}{2mc}\hat{\bm{A}}(\bm{r})\hat{\sigma}(\bm{r}).
\end{equation}
is the {\it effective interaction} current. The second and the third partitions of the system Hamiltonian $\hat{H}$ in Eq.~(\ref{ht}) involve the matter Hamiltonian $\hat{H}_{{\rm m}}$, that includes the kinetic energy of the electrons and their interactions with the transverse electromagnetic field; $\hat{H}_{{\rm c}}$ represents the Coulomb interactions between electrons that, as is well known \cite{cho2003optical}, can be interpreted as the energy of the longitudinal electromagnetic field (in particular the longitudinal field in the Coulomb gauge is a purely material variable), so that $\hat{H}_{{\rm f}}$ represents the total energy of the electromagnetic field.

Gauge invariance is most conveniently formulated using the electron field creation $\hat{\psi}^{\dagger}(\bm{r})$ and annihilation $\hat{\psi}(\bm{r})$ operators which satisfy the Fermi commutation relations
\begin{eqnarray}
\label{commutation-fermions} \{\hat{\psi}(\bm{r}),\hat{\psi}^{\dagger}(\bm{r}')\}= \delta(\bm{r}- \bm{r}')
\end{eqnarray}
the ``naive'', i.e., in the absence of the electromagnetic field, current $\hat{\bm{j}}(\bm{r})$ and charge density $\hat{\sigma}(\bm{r})$ operators are given by
\begin{eqnarray}
\label{jr}
\hat{\bm{j}}(\bm{r}) &=& \frac{e\hbar}{2mi}\left(\hat{\psi}^{\dagger}(\bm{r})\bm{\nabla}\hat{\psi}(\bm{r})- (\bm{\nabla}\hat{\psi}^{\dagger}(\bm{r}))\hat{\psi}(\bm{r})\right) \\
\label{rho}
\hat{\sigma}(\bm{r}) &=& e\hat{\psi}^{\dagger}(\bm{r})\hat{\psi}(\bm{r}).
\end{eqnarray}

All observables should be invariant to the following gauge transformation,
\begin{eqnarray}
\label{define-gauge-transform} &&\hat{\psi}(\bm{r}) \mapsto e^{i\varphi(\bm{r})}\hat{\psi}(\bm{r}), \;\;\; \hat{\psi}^{\dagger}(\bm{r}) \mapsto e^{-i\varphi(\bm{r})}\hat{\psi}^{\dagger}(\bm{r}), \nonumber \\
&& \hat{\bm{A}}(\bm{r}) \mapsto \hat{\bm{A}}(\bm{r})+ \frac{e}{c}\bm{\nabla}\varphi(\bm{r}).
\end{eqnarray}
A straightforward inspection shows that the following current $\hat{\bm{J}}(\bm{r})$ operator \cite{tanaka2001time},
\begin{eqnarray}
\label{int1} \hat{\bm{J}}(\bm{r})= \hat{\bm{j}}(\bm{r})-\frac{e^2}{mc}\hat{\bm{A}}(\bm{r})\hat{\sigma}(\bm{r}),
\end{eqnarray}
as well as the Hamiltonian $\hat{H}$ [Eq.~(\ref{ht})], with $\hat{H}_{{\rm int}}$ defined by Eq.~(\ref{define-H-int}),  [Eq.~(\ref{int1})], and the Hamiltonians $\hat{H}_{{\rm m}}$, $\hat{H}_{{\rm c}}$, $\hat{H}_{{\rm rad}}$, and $\hat{H}_{{\rm f}}$ are preserved by the gauge transformations. $\hat{\bm{J}}(\bm{r})$ will naturally appear below when calculating the matter/field energy exchange rate.

\section{The Matter/Field Energy Exchange Rate}
\label{sec:matter-field-exchange}

The optical response will be calculated via the matter-field energy exchange. To that end we define the energy of system $j$ as,
\begin{eqnarray}
\label{energy}
W_{j}(t)={\rm Tr}\left(\hat{H}_{j}\varrho(t)\right),
\end{eqnarray}
with $j= {\rm m}, {\rm c}, {\rm rad}, {\rm f}$, or without a subscript, stand for the energy of matter, Coulomb (longitudinal field), transverse field, total field, and total energy, respectively. Energy conservation implies$\dot{W}(t)\equiv dW(t)/dt= 0$. At equilibrium, $\dot{\varrho}(t)= 0$ and we further have $\dot{W}_{j}(t)= 0$ for all $j$, so that there is no energy exchange between the field and the matter. When the system is driven out of equilibrium, say by external fields or currents, the field-matter energy exchange rate becomes a measure of how strongly the external field drives the system out of its equilibrium. We will refer to $\dot{W}_{{\rm f}}= -\dot{W}_{{\rm m}}$ as the energy exchange rate between matter and field, where the field energy includes its longitudinal component. Another approach would be to consider the energy exchange rate $\dot{W}_{{\rm rad}}= -\dot{W}_{{\rm m}}- \dot{W}_{{\rm c}}$ between the transverse field and the system; in this second approach, the Coulomb energy, which is equal to the longitudinal field energy is included into the system.

The energy exchange rates can be evaluated by starting with Eq.~(\ref{energy})
\begin{eqnarray}
\label{energy-rates}
\dot{W}_{j}(t)&=&{\rm Tr}\left(\hat{H}_{j}\dot{\varrho}(t)\right)= -\frac{i}{\hbar}{\rm Tr}\left(\hat{H}_{j}[\hat{H},\varrho(t)]\right)\nonumber \\&=& -\frac{i}{\hbar}{\rm Tr}\left([\hat{H}_{j},\hat{H}]\varrho(t)]\right)= -\frac{i}{\hbar}\left\langle [\hat{H}_{j},\hat{H}]\right\rangle,
\end{eqnarray}
the commutators $[\hat{H}_{j},\hat{H}]$ are calculated using the electron operator commutation relations [Eq.~(\ref{commutation-fermions})], as well as the commutation relations between the field operators \cite{fetter2003quantum}. This gives,
\begin{eqnarray}
\label{energy-rates-explicit}
\dot{W}_{{\rm f}}(t)&=& -\frac{1}{2}\int d\bm{r}\left\langle \hat{\bm{E}}(\bm{r})\cdot \hat{\bm{J}}(\bm{r})+ \hat{\bm{J}}(\bm{r})\cdot \hat{\bm{E}}(\bm{r})\right\rangle\nonumber\\ &=& -\frac{1}{2}\int d\bm{r}{\rm Tr}\left(\left(\hat{\bm{E}}(\bm{r})\cdot \hat{\bm{J}}(\bm{r})+ \hat{\bm{J}}(\bm{r})\cdot \hat{\bm{E}}(\bm{r})\right)\varrho(t)\right),
\end{eqnarray}
where $\hat{\bm{E}}(\bm{r})= \hat{\bm{E}}_{{\rm T}}(\bm{r})+ \hat{\bm{E}}_{{\rm L}}(\bm{r})$, with $\hat{\bm{E}}_{{\rm L}}(\bm{r})$ being the (purely material) longitudinal electric field operator that represents the Coulomb field created by the charge density. The expressions for $\dot{W}_{{\rm c}}(t)$ and $\dot{W}_{{\rm rad}}(t)$ have a form of Eq.~(\ref{energy-rates-explicit}) with $\hat{\bm{E}}(\bm{r})$ replaced by $\hat{\bm{E}}_{{\rm L}}(\bm{r})$ and $\hat{\bm{E}}_{{\rm T}}(\bm{r})$, respectively.

Note that three different current operators appear naturally in the present description: (a) $\hat{\bm{j}}(\bm{r})$ [Eq.~(\ref{jr})] is the {\it naive} gauge-dependent current density operator and does not represent physical observable; (b) $\hat{\bm{J}}_{{\rm int}}$ [Eq.~(\ref{define-J-int})] is an effective interaction current operator that assures the gauge-invariance of the interaction Hamiltonian in Eq.~(\ref{define-H-int}); and (c) $\hat{\bm{J}}(\bm{r})$ [Eq.~(\ref{int1})] is the gauge-invariant current density operator which naturally appears in Eq.~(\ref{energy-rates-explicit}) when the energy exchange is calculated. Since the energy exchange rate is an observable, and therefore must be expressed in terms of gauge-invariant quantities.

We next consider the field-matter energy exchange rate in a system driven by and external field/current within the semiclassical approximation which neglects spontaneous processes. This boils down to replacing in Eq.~(\ref{energy-rates-explicit}) the electric field $\hat{\bm{E}}(\bm{r})$ and the vector potential $\hat{\bm{A}}(\bm{r})$, that enters $\hat{\bm{J}}(\bm{r})$ via Eq.~(\ref{int1}), with their classical counterparts, represented by the external (driving) field. This results in
\begin{eqnarray}
\label{absorptionrate} \dot{W}_{{\rm f}}(t)= -\int d\bm{r}\bm{E}(\bm{r},t)\cdot \bm{J}(\bm{r},t)
\end{eqnarray}

where,
\begin{eqnarray}
\label{define-J-average} \bm{J}(\bm{r},t)= \left\langle \hat{\bm{J}}(\bm{r},t) \right\rangle= {\rm Tr}\left(\hat{\bm{J}}(\bm{r})\hat{\varrho}(t)\right)
\end{eqnarray}
is the expectation value of the gauge-invariant current density, evaluated at the density matrix $\hat{\varrho}(t)$ of the driven system. The formal arguments in support of the intuitively natural procedure, we are using in this manuscript, as well as a way of calculating the radiative corrections will be addressed elsewhere.

It follows from Eqs.~(\ref{absorptionrate}) and (\ref{define-J-average}) that the semiclassical matter-field energy exchange rate for a driven system may be expressed in terms of the expectation value of the gauge-invariant current density $\bm{J}(\bm{r},t)$. It is thus natural to define the optical response functions as the expansion of the latter in powers of the driving field.

This can be represented in a compact form by introducing the following notation for Liouville space (tetradic) operators. With any Hilbert space operator $\hat{Q}$ we associate the following four Liouville space operators: {\it left} $\hat{Q}^{{\rm L}}$, {\it right} $\hat{Q}^{{\rm R}}$, {\it plus} $Q^{+}$, and {\it minus} $Q^{-}$, defined by its' action on another operator $\hat{X} $ \cite{marx2008nonlinear},
\begin{eqnarray}
\label{left right} &&\hat{Q}^{{\rm L}}(\hat{X}) \equiv \hat{Q}\hat{X}, \qquad \; \hat{Q}^{{\rm R}}(\hat{X}) \equiv \hat{X}\hat{Q}, \nonumber\\&& Q^{+} \equiv \frac{1}{2}(\hat{Q}^{{\rm L}}+ \hat{Q}^{{\rm R}}), \qquad \; \hat{Q}^{-}= \hat{Q}^{{\rm L}}- \hat{Q}^{{\rm R}}.
\end{eqnarray}
Adopting the interaction picture (with time parameter $\tau$) in Liouville space, we arrive at

\begin{eqnarray}
\label{J-average-semicl} \bm{J}(\bm{r},t;\bm{A};A_{0}) &=& \bigg\langle \bigg(\hat{\bm{j}}^{+}(\bm{r},t)- \frac{e}{mc}\bm{A}(\bm{r},t)\hat{\sigma}^{+}(\bm{r},t)\bigg)\nonumber\\&\times&  {\rm T}\exp\bigg(-i\hbar^{-1}\int_{-\infty}^{t}d\tau \hat{H}_{{\rm int}}^{-}(\bm{A},A_{0};\tau)\bigg) \bigg\rangle\nonumber\\ \end{eqnarray}
\begin{eqnarray}
\label{H-int-semicl} \hat{H}_{{\rm int}}^{-}(\bm{A},A_{0};\tau) &=& -\int d\bm{r}\bigg(\bm{A}(\bm{r})\cdot \hat{\bm{j}}^{-}(\bm{r},\tau) -\frac{e}{2mc}\bm{A}^{2}(\bm{r})\hat{\sigma}^{-}(\bm{r},\tau)\nonumber\\&+& A_{0}(\bm{r})\hat{\sigma}^{-}(\bm{r},\tau)\bigg),
\end{eqnarray}
the third term in the r.h.s. of Eq.~(\ref{J-average-semicl}) describes the interaction of the system's charge density with the longitudinal component of the driving field; in the Coulomb gauge, adopted here, it originates from the Coulomb interaction between the system and external charge density, the latter together with the external current density is responsible for creating the external (driving) field. The interaction Hamiltonian in Eq.~\ref{define-H-int} does not have a scalar potential component, since the contribution of the scalar potential is included in the $H_c$ of Eq.~\ref{ht}. The reason of explicitly including the scalar potential in Eq.~\ref{H-int-semicl} will be explained below.

Eq.~(\ref{absorptionrate}) together with Eqs.~(\ref{J-average-semicl}), and (\ref{H-int-semicl}) constitute the main formal result of this paper. A crucial feature of $\bm{J}(\bm{r},t;\bm{A};A_{0})$ [Eq.~(\ref{J-average-semicl})] is it's invariance with respect to time-dependent gauge transformations of the potentials
\begin{eqnarray}
\label{gauge-transform-time} \bm{A}(\bm{r},t) \mapsto \bm{A}(\bm{r},t)+ \frac{e}{c}\bm{\nabla}\varphi(\bm{r},t), \; A_{0}(\bm{r},t) \mapsto A_{0}(\bm{r},t)+ \frac{e}{c}\frac{\partial\varphi(\bm{r},t)}{\partial t}.
\end{eqnarray}
This property can be rationalized by noting that the transformation [Eq.~(\ref{gauge-transform-time})] in Eq.~(\ref{J-average-semicl}) is equivalent to performing the gauge transformation to the fermion fields in the interaction picture, which is given by Eq.~(\ref{define-gauge-transform}), but with a time-dependent $\varphi(\bm{r},\tau)$. This can be verified directly. The gauge invariance of the current $\bm{J}(\bm{r},t;\bm{A};A_{0})$ then follows from the fact that the gauge transformation applied to the fermion operators does not change their commutation relations. Also, the gauge invariance of the expectation value of the gauge-invariant current is intuitively clear, since any legitimate, i.e., gauge-invariant, observable should not change upon a gauge transformation of the external (driving) field.

Gauge invariance considerably simplifies the expressions for the optical response functions, presented below. Note that Eq.~(\ref{J-average-semicl}) has been derived, in the Coulomb gauge, so that the third term in the r.h.s. of the expression for $\hat{H}_{{\rm int}}$ [Eq.~(\ref{H-int-semicl})] is absolutely necessary. However, the gauge invariance of the optical response allows to use any gauge; for the sake of minimizing the number of terms in the expressions for the response functions we adopt the gauge $A_{0}(\bm{r})=0$. Then the third term in the expression for $\hat{H}_{{\rm int}}$ should be dropped whereas the transverse vector potential should be replaced with gauge invariant potential $\bm{A}^{{\rm inv}}(\bm{r},t)$,
\begin{eqnarray}
\label{A-gauge-inv} \bm{A}^{{\rm inv}}(\bm{r},t)= \int_{-\infty}^{t}d\tau \bm{E}(\bm{r},\tau)= \bm{A}(\bm{r},t)- \int_{-\infty}^{t}d\tau\bm{\nabla}A_{0}(\bm{r},\tau),
\end{eqnarray}
since, in the $A_{0}(\bm{r})= 0$ gauge we have $\dot{\bm{A}}= \bm{E}$. Note that the $A_{0}(\bm{r})= 0$ gauge provides an explicitly gauge-invariant picture of the optical response, which is expressed it in terms of the external electric field, or equivalently $\bm{A}^{{\rm inv}}$, both are gauge invariant.

The optical response functions are defined by an expansion of the current $\bm{J}(\bm{r},t;\bm{A};A_{0})= \bm{J}(\bm{r},t;\bm{A}^{{\rm inv}};0)$ in powers of $n$ driving field $\bm{A}^{{\rm inv}}$
\begin{eqnarray}
\label{define-response-funct} J_{k}(\bm{r},t) &=& \sum_{k_{1}}\int_{-\infty}^{t}d\tau_{1}\int d\bm{r}_{1}\zeta_{kk_{1}}^{(1)}(\bm{r},t;\bm{r}_{1},\tau_{1})A_{k_{1}}^{{\rm inv}}(\bm{r}_{1},\tau_{1}) \nonumber \\ &+& \sum_{n=2}^{\infty}\frac{1}{n!}\sum_{k_{1}\ldots k_{n}}\int d\tau_{1}d\bm{r}_{1}\ldots d\tau_{n}d\bm{r}_{n}\nonumber\\&\times& \zeta_{kk_{1}\ldots k_{n}}^{(n)}(\bm{r},t;\bm{r}_{1},\tau_{1},\ldots \bm{r}_{n},\tau_{n})A_{k_{1}}^{{\rm inv}}(\bm{r}_{1},\tau_{1})\ldots A_{k_{n}}^{{\rm inv}}(\bm{r}_{1},\tau_{n}),\nonumber\\
\end{eqnarray}
where $k,k_{1},\ldots,k_{n}= 1,2,3$.

The total energy exchange evaluated at time $t$ in Eq.~(\ref{absorptionrate}) can now be expanded order by order in driving fields. Note that time ordering is built into the response functions $\zeta^{(n)}(\bm{r},t;\bm{r}_{n},\tau_{n}...;\bm{r}_{1},\tau_{1})$. The first order energy loss rate is calculated  using Eqs.~(\ref{absorptionrate})$-$(\ref{define-response-funct}),
\begin{eqnarray}
\label{first order energy change}
\Delta \dot{W}^{(1)} (t)&=& -\sum_{kk_1}\int d\bm{r}  \int d\bm r_1 \int_{-\infty}^{t} d\tau_1 E_{k}(\bm{r},t) A_{k_{1}}^{{\rm inv}}(\bm{r}_1,\tau_1)\nonumber\\ &\times&\zeta_{kk_{1}}^{(1)}(\bm{r},t;\bm{r}_{1},\tau_{1}).
\end{eqnarray}

The first-order {\it non-local} response function is given by, 
\begin{eqnarray}
\label{zeta-1-expression-noscalar}
\zeta_{kk_1}^{(1)}(\bm{r},t;\bm{r}_{1},\tau_{1})&=& -i\hbar^{-1}\left\langle \hat{j}_{k}^{+}(\bm{r},t)\hat{j}_{k_{1}}^{-}(\bm{r}_{1},\tau_{1}) \right\rangle \nonumber\\&-& \frac{e}{mc}\delta_{kk_{1}}\left\langle \hat{\sigma}^{+}(\bm{r},t)\right\rangle \delta(t- \tau_{1})\delta(\bm{r}- \bm{r}_{1}).
\end{eqnarray}
This can be recast using ordinary Hilbert space operators as,
\begin{eqnarray}
\label{zeta-1-expression-hl} \zeta_{kk_1}^{(1)}(\bm{r},t;\bm{r}_{1},\tau_{1})&=& -i\hbar^{-1}\left\langle [\hat{j}_{k}(\bm{r},t),\hat{j}_{k_{1}}(\bm{r}_{1},\tau_{1})] \right\rangle \nonumber\\&-& \frac{e}{mc}\delta_{kk_{1}}\left\langle \hat{\sigma}(\bm{r},t)\right\rangle \delta(t- \tau_{1})\delta(\bm{r}- \bm{r}_{1}).
\end{eqnarray}

Superoperators are a convenient book-keeping device but at the end we can switch to ordinary operators. Now we turn to the non-linear matter/field energy exchange. Expanding Eq.~(\ref{absorptionrate}) to second order gives,
\begin{eqnarray}
\label{secondorderec}
\Delta \dot{W}^{(2)} (t) &=& -\sum_{k,k_1k_2}\int d\bm{r}  \int d\bm r_1 \int d\bm r_2 \int_{-\infty}^{t} d\tau_1 \int_{-\infty}^{t} d \tau_2 E_{k}(\bm{r},t)\nonumber\\&\times&A^{{\rm inv}}_{k_1}(\bm{r}_1,\tau_1) A^{{\rm inv}}_{k_2}(\bm{r}_2,\tau_2)\zeta_{kk_1k_1}^{(2)}(\bm{r}t;\bm{r}_{1}\tau_{1},\bm{r}_{2}\tau_{2})
\end{eqnarray}
where, the second order non-local response function, $\zeta_{kk_2k_1}^{(2)}(\bm{r}t;\bm{r}_2\tau_{2},\bm{r}_{1}\tau_{1})$ is,
\begin{eqnarray}
\label{zeta-2-expression}
&=& -\frac{1}{2}(i\hbar^{-1})^2\left\langle \hat{j}_{k}^{+}(\bm{r},t)\hat{j}_{k_2}^{-}(\bm{r}_2,\tau_2)\hat{j}_{k_1}^{-}(\bm{r}_1,\tau_1)\right\rangle\nonumber \\ &+& i\hbar^{-1}\frac{e}{2mc} \bigg( \delta_{k_1k_2}\left\langle \hat{j}_{k}^{+}(\bm{r},t)\hat{\sigma}^{-}(\bm{r}_1,\tau_1) \right \rangle\nonumber\\ &&\delta(\bm{r}_2-\bm{r}_1)\delta(\tau_2-\tau_1) + 2\delta_{kk_1}\delta(\bm{r}-\bm{r}_1) \nonumber\\ &&\delta(t-\tau_1)\left\langle\hat{\sigma}^{+}(\bm{r},t) \hat{\sigma}^{-}(\bm{r}_1,\tau_1)\right \rangle \bigg ),
\end{eqnarray}
which in Hilbert space becomes,
\begin{eqnarray}
\label{zeta-2-expression-hs}
&=& -\frac{1}{2}(i\hbar^{-1})^2\left\langle [[\hat{j}_{k}(\bm{r},t),\hat{j}_{k_2}(\bm{r}_2,\tau_2)],\hat{j}_{k_1}(\bm{r}_1,\tau_1)]\right\rangle\nonumber \\ &+& i\hbar^{-1}\frac{e}{2mc} \bigg( \delta_{k_1k_2}\left\langle [\hat{j}_{k}(\bm{r},t),\hat{\sigma}(\bm{r}_1,\tau_1)] \right \rangle\nonumber \\ && \delta(\bm{r}_2-\bm{r}_1)\delta(\tau_2-\tau_1) +  2\delta_{kk_1}\delta(\bm{r}-\bm{r}_1) \nonumber\\&&\delta(t-\tau_1)\left\langle[\hat{\sigma}(\bm{r},t), \hat{\sigma}(\bm{r}_1,\tau_1)]\right \rangle \bigg ).
\end{eqnarray}
Finally, the third order energy energy exchange,
\begin{eqnarray}
\label{thirdorder}
\Delta \dot{W}^{(3)} (t) &=& -\sum_{k,k_1k_2k_3}\int d\bm{r}   \int d\bm r_1 \dots \int d\bm r_3 \int_{-\infty}^{t} d\tau_1 \dots \int_{-\infty}^{t} d \tau_3\nonumber \\
&\times &E_{k}(\bm{r},t) A^{{\rm inv}}_{k_1}(\bm{r}_1,\tau_1) A^{{\rm inv}}_{k_2}(\bm{r}_2,\tau_2)A^{{\rm inv}}_{k_3}(\bm{r}_3,\tau_3)\nonumber \\
&\times &\zeta_{kk_3k_2k_1}^{(3)}(\bm{r},t;\bm{r}_3\tau_3,\bm{r}_2\tau_{2},\bm{r}_{1}\tau_{1}).
\end{eqnarray}
 Where, the third order non-local response function, $\zeta_{kk_3k_2k_1}^{(3)}(\bm{r}t;\bm{r}_3\tau_3,\bm{r}_2\tau_{2},\bm{r}_{1}\tau_{1})$,  is given by, 
\begin{eqnarray}
\label{zeta-three-nonlocal}
&=&-\frac{1}{6}(i\hbar^{-1})^3 \bigg\langle \hat{j}_{k}^{+}(\bm{r},t)\hat{j}_{k_3}^{-}(\bm{r}_3,\tau_3)\hat{j}_{k_2}^{-}(\bm{r}_2,\tau_2)\hat{j}_{k_1}^{-}(\bm{r}_1,\tau_1) \bigg \rangle \nonumber \\ &+&(i\hbar^{-1})^2\frac{e}{4mc} \bigg\{ \delta_{k_3k_2}
 \bigg\langle \hat{j}_{k}^{+}(\bm{r},t)\hat{j}_{k_2}^{-}(\bm{r}_2,\tau_2)\hat{\sigma}^{-}(\bm{r}_1,\tau_1)\bigg \rangle\nonumber\\&& \delta(\bm{r}_3-\bm{r}_2) \delta(\tau_3-\tau_2) \nonumber \\ &+& \delta_{k_3k_2}\left\langle \hat{j}_{k}^{+}(\bm{r},t)\hat{\sigma}^{-}(\bm{r}_2,\tau_2)\hat{j}_{k_1}^{-}(\bm{r}_1,\tau_1)\right \rangle \delta(\bm{r}_2-\bm{r}_1) \delta(\tau_2-\tau_1) \nonumber \\ &+& 2\delta_{kk_1}\left\langle \hat{\sigma}^{+}(\bm{r},t)\hat{j}_{k_2}^{-}(\bm{r}_2,\tau_2)\hat{j}_{k_1}^{-}(\bm{r}_1,\tau_1)\right \rangle \delta(\bm{r}-\bm{r}_1) \delta(t-\tau_1) \bigg\} \nonumber \\ &-& (i \hbar) \frac{e^2}{2mc^2}\delta_{kk_1} \left \langle \hat{\sigma}^{+}(\bm{r},t)\hat{\sigma}^{-}(\bm{r}_1,\tau_1)\right \rangle \delta(\bm{r}_3-\bm{r}_2)\nonumber \\&& \delta(\bm{r}-\bm{r}_1)  \delta(\tau_3-\tau_2)\delta(t-\tau_1).
\end{eqnarray}  
In Hilbert space it is given by,
\begin{eqnarray}
\label{zeta-three-nonlocal-hs}
&=&-\frac{1}{6}(i\hbar^{-1})^3 \bigg\langle[[[\hat{j}_{k}(\bm{r},t),\hat{j}_{k_3}(\bm{r}_3,\tau_3)],\hat{j}_{k_2}(\bm{r}_2,\tau_2)],\hat{j}_{k_1}(\bm{r}_1,\tau_1)] \bigg \rangle \nonumber \\ &+&(i\hbar^{-1})^2\frac{e}{4mc} \bigg\{ \delta_{k_3k_2}
 \bigg\langle [[\hat{j}_{k}(\bm{r},t),\hat{j}_{k_2}(\bm{r}_2,\tau_2)],\hat{\sigma}(\bm{r}_1,\tau_1)]\bigg \rangle \nonumber\\ &&\delta(\bm{r}_3-\bm{r}_2) \delta(\tau_3-\tau_2) \nonumber \\ &+& \delta_{k_3k_2}\left\langle [[\hat{j}_{k}(\bm{r},t),\hat{\sigma}(\bm{r}_2,\tau_2)],\hat{j}_{k_1}(\bm{r}_1,\tau_1)]\right \rangle \delta(\bm{r}_2-\bm{r}_1) \delta(\tau_2-\tau_1) \nonumber \\ &+& 2\delta_{kk_1}\left\langle [[\hat{\sigma}^{+}(\bm{r},t),\hat{j}_{k_2}(\bm{r}_2,\tau_2)],\hat{j}_{k_1}(\bm{r}_1,\tau_1)]\right \rangle \delta(\bm{r}-\bm{r}_1) \delta(t-\tau_1) \bigg\} \nonumber \\ &-& (i \hbar) \frac{e^2}{2mc^2}\delta_{kk_1} \left \langle [\hat{\sigma}(\bm{r},t),\hat{\sigma}(\bm{r}_1,\tau_1)]\right \rangle\nonumber\\&& \delta(\bm{r}_3-\bm{r}_2)\delta(\bm{r}-\bm{r}_1)  \delta(\tau_3-\tau_2)\delta(t-\tau_1).
\end{eqnarray}
Eqs.~(\ref{zeta-1-expression-hl}),(\ref{zeta-2-expression-hs}) and (\ref{zeta-three-nonlocal-hs}) constitute our final exact expressions for the non-local response functions, given by correlation functions of the charge density ($\hat{\sigma}(\bm r,t)$) and the current density operator ($\bm{\hat{j}}(\bm r,t) $). The current and charge densities in Eqs.~(\ref{zeta-1-expression-hl}),(\ref{zeta-2-expression-hs}) and (\ref{zeta-three-nonlocal-hs}) can be computed using the standard output of quantum chemistry packages, like Molpro \cite{MOLPRO}, by calculating the relevant molecular orbitals, then explicitly evaluating transition current and charge densities Eqs.~(\ref{jr} and \ref{rho}), and, expressing them in terms of superoperators using Eqs.~(\ref{left right}). Note that in the non-local representation magnetic and electric contributions need not be treated separately which greatly simplifies the analysis. 

\section{Recovering the Dipole Approximation}
\label{sec:dipole app}

Formally, the minimal coupling and the multipolar Hamiltonians are connected by the Power-Zienau transformation in the joint space of matter and field degrees of freedom \cite{power1959philos}.  However, some discrepancies arise when the two Hamiltonians are applied in practice. A known example is an extra pre-factor of $(\omega_{\alpha}/\Omega)^2$ with material frequency ($\omega_{\alpha}$) and field frequency ($\Omega$) in the absorption lineshape \cite{lamb1987matter} [See Appendix A]. The issue was resolved by making the canonical transformation between the two Hamiltonians. The non-locality of the response implies that a field acting at one point includes a charge or current at a different point, suggesting an induced electronic coherence. While the multipolar expansion is computationally convenient, it hides this interesting non-local physical picture. The non-local response functions avoid the tedious canonical transformation to the multipolar Hamiltonian \cite{lamb1987matter}. And show no discrepancy between the dipole and minimal coupling absorption lineshapes. 

As a consistency check, here we show how the dipole approximation for the linear response is recovered for the nonlinear minimal coupling response in agreement with the multipolar Hamiltonian. We consider a system of the size, small compared to the field wavelength, which allows us to drop the $\bm{r}$-dependence in the fields $\bm{E}$ and $\bm{A}^{\rm{inv}}$ and recast Eq.~(\ref{first order energy change}) in a form
\begin{eqnarray}
\label{W-lin-dipole} \Delta\dot{W}_{\rm{d}}^{(1)}= i\hbar^{-1}\bm{E}(t)\cdot \bm{u}(t)+\frac{Ne^{2}}{mc}\bm{E}(t)\cdot \bm{A}^{\rm{inv}}(t),
\end{eqnarray}
where $N$ is the number of electrons,
\begin{eqnarray}
\label{define-u} \bm{u}(t)= \sum_{k,s}\int_{-\infty}^{t}dt'\left\langle \hat{u}_{k}^{+}(t)\hat{u}_{s}^{-}(t') \right\rangle A_{s}^{\rm{inv}}(t'),
\end{eqnarray}
$k,s= 1,2,3$ and
\begin{eqnarray}
\label{define-hat-u} \hat{\bm{u}}(t)= \int d\bm{r}\hat{\bm{j}}(\bm{r},t)= -\int d\bm{r}\bm{r}\left(\bm{\nabla} \cdot \hat{\bm{j}}(\bm{r},t)\right)= \frac{d\hat{\bm{\mu}}(t)}{dt},
\end{eqnarray}
with $\hat{\bm{\mu}}$ being the dipole operator. The above expressions can be interpreted as follows: the operator $\hat{\bm{u}}$ is the integral over the space of the naive (i.e., in the absence of the field) current density operator, the second equality in Eq.~(\ref{define-hat-u}) is due to the Stokes formula, whereas the last equality follows from the continuity equation
\begin{eqnarray}
\label{continuity-eq} \bm{\nabla} \cdot \hat{\bm{j}}(\bm{r},t) + \frac{d\hat{\sigma}(\bm{r},t)}{dt} = 0
\end{eqnarray}
in the non-driven system that holds on the operator level, combined with the well-known definition of the dipole operator $\hat{\bm{\mu}}$. As a result $\hat{\bm{u}}$ can be viewed as the time derivative of the dipole. The variable $\bm{u}(t)$ is the expectation value of $\hat{\bm{u}}(t)$ in the linearly driven system; Eq.~(\ref{W-lin-dipole}) is obtained by integration over $\bm{r}$ and $\bm{r}'$ in Eq.~(\ref{first order energy change}) under the assumption that the fields do not depend on $\bm{r}$.

Substituting the expression for $\hat{\bm{u}}(t)$ as the time derivative of the dipole [Eq.~(\ref{define-hat-u})] into Eq.~(\ref{define-u}), followed by substituting the latter into Eq.~(\ref{W-lin-dipole}), and further integrating over $t'$ by parts, as well as taking the derivative with respect to $t$ out of the integral over $t'$ we arrive at
\begin{eqnarray}
\label{W-lin-dipole-2} \Delta\dot{W}_{\rm{d}}^{(1)} &=& -i\hbar^{-1}\sum_{ks}E_{k}(t)\frac{d}{dt}\int_{-\infty}^{t}dt'\left\langle \hat{\mu}_{k}^{+}(t)\hat{\mu}_{s}^{-}(t') \right\rangle \dot{A}_{s}^{\rm{inv}}(t') \nonumber \\ &+& i\hbar^{-1}\sum_{ks}E_{k}(t)\frac{d}{dt}\left(\left\langle \hat{\mu}_{k}^{+}(t)\hat{\mu}_{s}^{-}(t-0) \right\rangle\right) A_{s}^{\rm{inv}}(t) \nonumber \\ &-& i\hbar^{-1}\sum_{ks}E_{k}(t)\left\langle \hat{\mu}_{k}^{+}(t)\hat{u}_{s}^{-}(t-0) \right\rangle A_{s}^{\rm{inv}}(t) \nonumber \\ &+& \frac{Ne^{2}}{mc}\bm{E}(t)\cdot \bm{A}^{\rm{inv}}(t),
\end{eqnarray}
where the second term in the r.h.s. of Eq.~(\ref{W-lin-dipole-2}) is a contact (boundary) contribution in the by parts integration over $t'$, whereas the third term compensates for the contact (boundary contribution) that occurs in the time derivative with respect to $t$ of an integral, whose upper limit is $t$-dependant. These two terms can be computed using the fact that for any two operators $\hat{X}$ and $\hat{Y}$, we have $\langle \hat{X}^{+}(t)\hat{Y}^{-}(t-0) \rangle = \langle [\hat{X}(t),\hat{Y}(t)] \rangle$. This implies that the first contact term vanishes, whereas the second one is obtained using the well-known commutation relation (that can also be verified directly using a simple and straightforward calculation)
\begin{eqnarray}
\label{j-rho-commutator} [\hat{j}_{k}(\bm{r}'),\hat{\sigma}(\bm{r})] = \frac{ie^{2}\hbar}{mc}\frac{\partial}{\partial r_{k}}\delta(\bm{r} - \bm{r}').
\end{eqnarray}
After some simple and straightforward transformations, we obtain, 
\begin{eqnarray}
\label{mu-u-commutator} [\hat{\mu}_{k},\hat{u}_{s}] = -\frac{i\hbar Ne^{2}}{mc}\delta_{ks}
\end{eqnarray}
The third and fourth terms in the r.h.s. of Eq.~(\ref{W-lin-dipole-2}) now cancel each other. Recalling the standard definition of the time-domain linear response function in the dipole approximation
\begin{eqnarray}
\label{define-chi-lin-dipole} \chi_{ks}^{(1)}(t) = i\hbar^{-1}\left\langle \hat{\mu}_{k}^{+}(t)\hat{\mu}_{s}^{-}(t) \right\rangle,
\end{eqnarray}
and the fact that $\dot{\bm{A}}^{\rm{inv}}(t) = \bm{E}(t)$, we arrive at
\begin{eqnarray}
\label{W-lin-dipole-3} \Delta\dot{W}_{\rm{d}}^{(1)} = -\sum_{ks}E_{k}(t)\frac{d}{dt}\int_{-\infty}^{t}dt'\chi_{ks}^{(1)}(t-t')E_{s}(t')
\end{eqnarray}
that recovers the dipole approximation for the energy exchange rate in terms of the standard response function $\chi_{ks}^{(1)}(t)$.

We first note that, the simple expression for the dipole approximation [Eq.~(\ref{W-lin-dipole-3})] was obtained by the cancellation of the last two terms in Eq.~(\ref{W-lin-dipole-2}), which would be impossible if the second (local) term in the expression for the linear response [Eq.~(\ref{first order energy change})] was omitted. The derivation, presented in this section thus does not show the  ``extra pre-factor of $(\omega/\Omega)^2$" in absorption spectra \cite{craig84,lamb1987matter}. Second, if we define
\begin{eqnarray}
\label{P-J-mu} \hat{\mu} = \int d\bm{r}\hat{\bm{P}}(\bm{r}), \;\;\; \hat{\bm{J}}(\bm{r},t)= - \frac{d\hat{\bm{P}}(\bm{r},t)}{dt},
\end{eqnarray} 
as is done in the multipolar Hamiltonian formalism with the magnetic terms neglected, one immediately obtains from Eq.~(\ref{W-lin-dipole-3})
\begin{eqnarray}
\label{W-lin-dipole-4} \Delta\dot{W}_{\rm{d}}^{(1)} = -\bm{E}(t) \cdot \int d\bm{r}\bm{J}(\bm{r},t), 
\end{eqnarray}
thereby recovering the long wavelength limit of Eq.~(\ref{absorptionrate}). Third, an attempt to compute the dipole approximation for the energy exchange rate, starting with the ``standard'' dipole approximation Hamiltonian
\begin{eqnarray}
\label{H-int-dipole} \hat{H}^{{\rm d}}_{{\rm int}}=-\hat{\bm{\mu}}\cdot \bm{E}(t),
\end{eqnarray}
will result in an expression, different from Eq.~(\ref{W-lin-dipole-3}). This can be explained as follows. The dipole Hamiltonian [Eq.~(\ref{H-int-dipole})] is naturally obtained from the multipolar formalism that can be described as using a multipolar gauge which is different from the Coulomb gauge, or equivalently doing a canonical transformation that redistributes the energy between the matter and the field. Therefore, a calculation based on Eq.~(\ref{H-int-dipole}) addresses an exchange between energies, different from the ones, used in our calculations.

\section{Heterodyne Detected Signals}
Heterodyne detection measures the energy exchange with a given wavevector component of the field rather than the entire field. The heterodyne detected signal can be recovered from the energy exchange rate \cite{marx2008nonlinear}. We start with the (transverse) vector potential operator $\hat{\bm{A}}(\bm{r},t)$ expanded in modes $\bm{q}\lambda$ is,
\begin{eqnarray}
\label{A}
\hat{\bm{A}}(\bm{r}) = \sum_{\bm{q}\lambda}c\frac{\epsilon_{\bm{q}\lambda}}{\omega_{\bm{q}}}\bm{e}_{\bm{q}\lambda} \left(\hat{b}_{{\bm{q}\lambda}}^{\dagger}e^{-i\bm{q}\cdot \bm{r}}+ \hat{b}_{{\bm{q}\lambda}}e^{i\bm{q}\cdot \bm{r}}\right)
\end{eqnarray}
where $c$ is speed of light, $\epsilon_{\bm{q}\lambda}=(2\pi\hbar\omega_{\bm{q}}/V)^{1/2}$ and $\bm{e}_{\bm{q}\lambda}$ is the polarization vector, while $V$ is the quantization volume. The transverse nature of the photon field manifest itself in the relation $\bm{q}\cdot \bm{e}_{\bm{q}\lambda}= 0$.

The heterodyne detected signal $\mathcal{S}_{{\rm het}} $ can be defined as the rate of change of the photon number $\hat{N}_{s}= \hat{b}^{\dagger}_{\bm{q}_{s}\lambda_{s}} \hat{b}_{\bm{q}_{s}\lambda_{s}}$ in mode $s$ \cite{marx2008nonlinear},
\begin{eqnarray}
\label{shet-cm}
\mathcal{S}_{{\rm het}}(s)= \frac{d \langle \hat{N}_{s}  \rangle}{dt} = \frac{i}{\hbar}\left[\hat{H},\hat{N}_{s}\right]
\end{eqnarray}
where the expectation value $\langle \hat{N}_{s}  \rangle $ is taken with respect to material density matrix. More precisely, the heterodyne detected signal is the cross-component obtained by mixing the signal field (i.e., generated by the currents induced in the material by the driving field) with the heterodyne counterpart. Evaluating the commutations gives,
\begin{equation}
\label{het}
\mathcal{S}_{{\rm het}}(s)= \frac{dB_{\bm{q}_{s}\lambda_{s}}^{*}}{dt}b_{\bm{q}_{s}\lambda_{s}}+ B_{\bm{q}_{s}\lambda_{s}}^{*}\dot{b}_{\bm{q}_{s}\lambda_{s}}+ {\rm c.c.} \;\;\; \dot{b}_{\bm{q}_{s}\lambda_{s}}= \frac{db_{\bm{q}_{s}\lambda_{s}}}{dt}
\end{equation}
with
\begin{eqnarray}
\label{define-averages} B_{\bm{q}_{s}\lambda_{s}}= \left\langle \hat{b}_{\bm{q}_{s}\lambda_{s}}\right\rangle_{{\rm het}}, \;\;\; b_{\bm{q}_{s}\lambda_{s}}= \left\langle \hat{b}_{\bm{q}_{s}\lambda_{s}} \right\rangle= {\rm Tr}\left(\hat{b}_{\bm{q}_{s}\lambda_{s}}\hat{\varrho(t)}\right),\nonumber\\
\end{eqnarray}
so that $B_{\bm{q}_{s}\lambda_{s}}(t)= \bar{B}_{\bm{q}_{s}\lambda_{s}}(t)e^{-i\omega_{\bm{q}_{s}}t}$ is a classical field that represents the heterodyne and is linear combination of gauge-invariant electric and transverse vector potential with temporal envelops $(\bar{E}(t))$ and $(\bar{A}(t)) $ at the same mode $\bm {q}_s$, i.e., $B_{\bm{q}_{s}\lambda_{s}}(t)= (\bar{E}_{\bm{q}_{s}\lambda_{s}}(t)\pm \bar{A}_{\bm{q}_{s}\lambda_{s}}(t))e^{-i\omega_{\bm{q}_{s}}t}$ where $\bar{A}_{\bm{q}_{s}\lambda_{s}}(t)$ is the temporal envelop of vector potential. Whereas $b_{\bm{q}_{s}\lambda_{s}}(t)$ is the expectation value of the photon annihilation operator, evaluated at the density matrix $\hat{\varrho}(t)$ of the driven system. The latter can be evaluated using the Heisenberg equation of motion
\begin{eqnarray}
\label{b-evolution} \dot{b}_{\bm{q}_{s}\lambda_{s}}&=& -\frac{i}{\hbar}\left\langle[\hat{H},\hat{b}_{\bm{q}_{s}\lambda_{s}}]\right\rangle\nonumber\\ &=& -i\omega_{\bm{q}_{s}}b_{\bm{q}_{s}\lambda_{s}}+ \frac{i}{\hbar}\int d\bm{r}\left\langle\hat{\bm{J}}(\bm{r})\cdot \left[\hat{b}_{\bm{q}_{s}\lambda_{s}},\hat{\bm{A}}(\bm{r})\right]\right\rangle \nonumber \\ &=& -i\omega_{\bm{q}_{s}}b_{\bm{q}_{s}\lambda_{s}} + \frac{ic\epsilon_{\bm{q}_{s}\lambda_{s}}}{\hbar\omega_{\bm{q}_{s}}}\int d\bm{r}e^{-i\bm{q}_{s}\cdot \bm{r}}\bm{e}_{\bm{q}_{s}\lambda_{s}}\cdot \bm{J}(\bm{r},t).\nonumber\\
\end{eqnarray}

The second equality is obtained by an explicit computation of the commutator, in $\bm{J}(\bm{r},t)$ as given in Eq.~(\ref{define-J-average}). Summary: It follows from Eqs.~(\ref{het}), (\ref{define-averages}), (\ref{b-evolution}), and (\ref{define-J-average}) that the heterodyne signal, introduced earlier in a fully quantum way and in terms of the field variables, can be expressed in terms of the expectation value of the gauge-invariant current density operator $\hat{\bm{J}}(\bm{r})$. This result is exact, and in particular can be considered as a starting point for calculating the radiative corrections to the signals. The latter, however goes beyond the scope of this letter, and will be addressed elsewhere.

The heterodyne signal has the form,
\begin{eqnarray}
\label{S-het-explicit} {\cal S}_{{\rm het}}(s) &=& (\dot{\bar{E}}_{\bm{q}_{s}\lambda_{s}}^{*}(t)\pm \dot{\bar{A}}_{\bm{q}_{s}\lambda_{s}}^{*}(t) )\nonumber\\ &&\frac{ic\epsilon_{\bm{q}_{s}\lambda_{s}}}{\hbar\omega_{\bm{q}_{s}}}\int_{-\infty}^{t}dt\int d\bm{r}e^{i\omega_{\bm{q}_{s}}t -i\bm{q}_{s}\cdot \bm{r}}\bm{e}_{\bm{q}_{s}\lambda_{s}}\cdot \bm{J}(\bm{r},t) \nonumber \\ &+& (\bar{E}_{\bm{q}_{s}\lambda_{s}}^{*}(t)\pm \bar{A}_{\bm{q}_{s}\lambda_{s}}^{*}(t) )\nonumber\\ && \frac{ic\epsilon_{\bm{q}_{s}\lambda_{s}}}{\hbar\omega_{\bm{q}_{s}}}\int d\bm{r}e^{i\omega_{\bm{q}_{s}}t -i\bm{q}_{s}\cdot \bm{r}}\bm{e}_{\bm{q}_{s}\lambda_{s}}\cdot \bm{J}(\bm{r},t)+ {\rm c.c.}\nonumber\\
\end{eqnarray}

Under the semi-classical approximation the interaction picture expansion of this expectation value in orders of incoming $n$ vector fields are done in Eqs.~(\ref{J-average-semicl}) $-$ (\ref{define-response-funct}). The heterodyne signal is thus given by the same response functions introduced for the energy exchange.

The above representation is gauge invariant, it allows to reduce the summations to purely spatial values, and, finally provides a connection to the ``standard'' picture of optical response. Introducing formally the polarization $\bm{P}$ by $\dot{\bm{P}}= -\bm{J}$, we can interpret Eq.~(\ref{J-average-semicl}), written in the $A_{0}=0$ gauge, and combined with Eq.~(\ref{A-gauge-inv}) as an expansion of the system polarization in powers of the driving electric field. This is possible, despite the fact that there is no simple polarization operator, and the material system interacts with the driving field via the scalar and vector potentials, rather than the electric field. Introducing the polarization is not necessary in the present formalism and the comment was made to connect with the standard formalism.

\section{Conclusions}
We have shown that physical observables such as matter-field energy exchange and the heterodyne detected optical signal in Eq.~(\ref{het}) depend on expectation value of the gauge-invariant current density operator $\hat{\bm{J}}(\bm{r},t;\bm{A})$. Note that $\sigma^{+}(\bm{r},t)$ in Eqs.~(\ref{zeta-1-expression-noscalar}), (\ref{zeta-2-expression}) and (\ref{zeta-three-nonlocal}) comes from the gauge-invariant current  (Eq.~(\ref{int1})) whereas $\sigma^{-}(\bm{r},t)$ comes from $H_{{\rm int}}$ in Eq.~(\ref{int}). We further presented a gauge-invariant non-local formal response function in Liouville space to any order [Eq.~(\ref{J-average-semicl})] and expanded it to first, second and third orders [Eqs.~(\ref{zeta-1-expression-noscalar}), (\ref{zeta-2-expression}) and (\ref{zeta-three-nonlocal})], together with the corresponding Hilbert space representations [Eqs.~(\ref{zeta-2-expression-hs}), (\ref{zeta-2-expression-hs}), and (\ref{zeta-three-nonlocal-hs})]. The multipolar and minimal-coupling matter-field interaction Hamiltonians are consistent as shown in Eqs.~(\ref{W-lin-dipole}) $-$ (\ref{W-lin-dipole-4}). The complexity of the multipolar formalism grows rapidly for the non-linear optical response when using both electric and magnetic multipoles in nano-shaped lights \cite{novotny2012principles}, which can be avoided in this formalism. The non-local response functions allow us to exactly calculate heterodyne detected optical signals in the presence of strong fields and non-uniform nano optical fields \cite{keller2012quantum,novotny2012principles}. The formalism can be used to study non-adiabatic molecular current density dynamics \cite{mandal2015non}. Furthermore, it can be extended to cavity quantum electrodynamics (QED), or be used in understanding field angular momentum \cite{heeres2014subwavelength}.

\section*{Acknowledge}
The authors gratefully acknowledge the support of National Science Foundation (Grants No. CHE-1361516 and CHE-1111350) and the Chemical Sciences, Geosciences and Biosciences Division, Office of Basic Energy Sciences, Office of Science, U.S. Department of Energy.

\appendix

\setcounter{section}{0}
\section{Lamb discrepancy of $(- i \omega_{\alpha 0}/\omega)^2$}

We present the frequency domain calculation of linear non-local field/matter energy exchange which recovers the discrepancy of $(-i \omega_{\alpha 0}/\omega)^2$ in the linear absorption spectra when using ``naive'' minimal coupling Hamiltonian and dipole approximation \cite{lamb1987matter}. In the dipole approximation the field/matter couplings is given by \cite{mukamel1995,scully1997},
\begin{eqnarray}
\label{Hintdip}
\hat{H}^d_{int} = - \bm {\hat{\mu}}\cdot \bm E(t),
\end{eqnarray}
where, $\bm {\hat{\mu}} $ is dipole operator, $\bm E(t)$ is the electric field; in the dipole approximation $\bm r$- dependence of the electric field is ignored. Thus, standard response theory gives us,
\begin{eqnarray}
\label{dipoleresponse}
\Delta W_{d}^{(1)}(t)= -i\hbar^{-1} \int_{-\infty}^{t} d\tau \dot{\bm E}(t)\cdot \left\langle\bm {\hat{\mu}}^{+}(t)\bm {\hat{\mu}}^{-}(\tau) \right\rangle\cdot \bm E(\tau)\nonumber\\
\end{eqnarray}
Or in frequency domain as,
\begin{eqnarray}
\label{dipoleenergyfre}
\Delta W ^{(1)}_{d} &=& \omega \int d\omega  |\bm E(\omega)|^2  \mathcal{S}_d(\omega),\nonumber\\
\mathcal{S}_d(\omega)&=& -\hbar^{-1}\int_{0}^{\infty} d\tau  \left\langle \bm \mu^{+}(\tau)\bm \mu^{-}(0) \right \rangle exp(i \omega \tau)
\end{eqnarray}
Expressing Eq.~(\ref{dipoleenergyfre}) in sum over states and using $\bm E(\omega)=2 \pi E_0 \delta(\omega-\omega') $, we can write,
\begin{eqnarray}
\label{dipolefreqsos}
\Delta W ^{(1)}_{d} &=& -2 |E_0|^2 \omega Im  \bigg\{\bigg(\frac{n_0}{\hbar}\sum_{\alpha}[P(\alpha)-P(0)]\nonumber\\&\times&\frac{|\mu_{\alpha 0}|^2}{\omega-\omega_{\alpha 0}+i\eta_{\alpha 0}}\bigg)\bigg\}\nonumber \\
\end{eqnarray}
Here, $n_0$ is the total number of particles per unit volume and $P(0),\; P(\alpha)$ are the thermal population of state $|0\rangle \; |\alpha\rangle$ respectively and are defined in usual way. $\mu_{\alpha 0}$ and $\omega_{\alpha 0}$ are relevant electric dipole elements and energy related to the electronic transition $|0\rangle \rightarrow |\alpha\rangle $.

A ``naive'' application of the minimal coupling Hamiltonian neglects the $\bm A^2(\bm r,t)$ since it is non-linear. This then gives,
\begin{eqnarray}
\label{mcenergy}
\Delta W ^{(1)}_{mc}(t) = -i \hbar^{-1} \int_{-\infty}^{t} d\tau_1 \bm \ddot{A}(t) \cdot\left\langle \hat{\bm j}^{+}(t)\hat{\bm j}^{-}(\tau_1) \right \rangle \cdot \bm A(\tau_1).\nonumber\\
\end{eqnarray}
Or in frequency domain as,
\begin{eqnarray}
\label{mceenergyfre}
\Delta W ^{(1)}_{mc} &=& \omega\int d\omega |\bm A(\omega)|^2  \mathcal{S}_{mc}(\omega),\nonumber\\
\mathcal{S}_{mc}(\omega)&=& -\hbar^{-1}\int_{0}^{\infty} d\tau \left\langle \hat{\bm j}^{+}(\tau)\hat{\bm j}^{-}(0) \right \rangle exp(i \omega \tau) \nonumber \\
&=& \bigg(\frac{n_0}{\hbar}\sum_{\alpha}[P(\alpha)-P(0)]\frac{ |j_{\alpha 0}|^2 }{\omega-\omega_{\alpha 0}+i\eta_{\alpha 0}}\bigg).
\end{eqnarray}
The current density elements for given transition, (say, $|0\rangle \rightarrow |\alpha\rangle $), $ j_{ 0 \alpha}= j^{*}_{ \alpha 0}=\langle \alpha|\hat{\bm j}|0\rangle$ is related to $\textit{transition dipole}$, $ \mu_{0 \alpha}(t)=\mu_{0 \alpha} exp(-i \omega_{0 \alpha} t)$ by,
\begin{eqnarray}
\label{jrelmu}
\mu_{0 \alpha} = -i (\omega_{0 \alpha})^{-1} j_{ 0 \alpha},
\end{eqnarray}
since $\bm E(\omega)= i \omega \bm A( \omega)$, the first order matter/radiation energy exchange in $\textit{minimal coupling}$ thus becomes,
\begin{eqnarray}
\label{energymcfreq}
\Delta W ^{(1)}_{mc}=\left(\frac{-i\omega_{\alpha 0}}{\omega}\right)^2\Delta W ^{(1)}_{d} .
\end{eqnarray}
As can be seen from Eqs.~(\ref{dipoleenergyfre})-(\ref{energymcfreq}), the matter/field energy exchange differ by a factor of $(-i\omega_{\alpha 0}/\omega)^2$ in the minimal coupling (ignoring $\bm A^2(\bm r,t)$) and dipole approximation. This factor adds asymmetry to the absorption peak and was resolved by \cite{lamb1987matter} using tedious canonical transformation. This apparent discrepancy is caused by the ''naive`` use of the minimal coupling Hamiltonian rather than gauge invariant interaction as shown in main text under section $\textit{Recovering Dipole Approximation}$.


\end{document}